\documentclass{article}
\usepackage{spconf,amsmath,graphicx}
\usepackage{amsfonts}
\usepackage{url}
\usepackage{float}
\usepackage{siunitx}

\usepackage[font=small,labelfont=bf]{caption}
\title{Content Based Singing Voice Extraction From a Musical Mixture}
\name{Pritish Chandna\textsuperscript{1}, Merlijn Blaauw\textsuperscript{1}, Jordi Bonada\textsuperscript{1}, Emilia G\'omez\textsuperscript{1,2}\thanks{The TITANX used for this research was donated by the NVIDIA Corporation. This work is partially supported by the Towards Richer Online Music Public-domain Archives (TROMPA H2020 770376) project. The dataset used for training was provided by Yamaha Corporation.}}
\address{\textsuperscript{1}Music Technology Group, Universitat Pompeu Fabra, Barcelona, Spain\\
\textsuperscript{2}Joint Research Centre, European Commission, Seville, Spain}
\begin{document}
\setlength{\abovedisplayskip}{4pt}
\setlength{\belowdisplayskip}{4pt}
\setlength{\abovedisplayshortskip}{4pt}
\setlength{\belowdisplayshortskip}{4pt}
%
\maketitle
\begin{abstract}
We present a deep learning based methodology for extracting the singing voice signal from a musical mixture based on the underlying linguistic content. Our model follows an encoder-decoder architecture and takes as input the magnitude component of the spectrogram of a musical mixture with vocals. The encoder part of the model is trained via knowledge distillation using a teacher network to learn a content embedding, which is decoded to generate the corresponding vocoder features.
Using this methodology, we are able to extract the unprocessed raw vocal signal from the mixture even for a processed mixture dataset with singers not seen during training. While the nature of our system makes it incongruous with traditional objective evaluation metrics, we use subjective evaluation via listening tests to compare the methodology to state-of-the-art deep learning based source separation algorithms. We also provide sound examples and source code for reproducibility. 
\end{abstract}
\section{Introduction}
\label{sec:intro}

Source separation, or the process of isolating individual signals from a mixture of such signals has been long studied, traditionally with statistical approaches like non-negative matrix factorization (NMF)~\cite{Durrieu2009}, principal component analysis~\cite{huang2012singing} and independent component analysis (ICA)~\cite{comon1994independent}. In the last few years, methodologies based on deep learning have been proposed, leading to a significant advance in the field. Source separation has diverse applications across fields; medical imaging, image processing and financial modelling among others. In music, the most useful application is that of separating the lead vocals from a musical mixture. This problem is well researched and numerous deep learning based models have recently been proposed to tackle it~\cite{huang2014singing, chandna2017monoaural, nugraha2016multichannel, jansson2017singing,venkataramani2019style, sahai2019spectrogram, luo2018tasnet, stoller2018wave}. Most of these models use the neural network to predict soft time frequency masks, given an input magnitude spectrogram of the mixture signal. This mask is then applied to the magnitude spectrogram to isolate the desired signal, which is re-synthesised by using the phase of the mixture spectrogram. The output of such algorithms is the processed vocal signal, sometimes referred to as the \textit{stem}~\cite{bittner2014medleydb}. 

The \textit{stem} is the raw audio signal with linear and non-linear effects such as reverb, compression, delay and equalization among others. We recently proposed a model~\cite{chandna2019vocoder}, which re-synthesises the underlying raw vocal signal present in a musical signal by using a deep neural network architecture as a function approximator that predicts vocoder features pertaining to the vocal signal. The model outperformed state-of-the-art source separation algorithms in terms of isolation of the signal from the backing track. 

In this paper, we present a methodology for synthesising raw, unprocessed vocals from a musical mixture based on human cognition; i.e., we first extract a representation of the linguistic content present in the mixture and then generate the vocal signal based on this content. The linguistic content pertains primarily to cognitively relevant features such as phonemes, however we do not explicitly predict the phonemes present in the signal. Although we did explore this possibility, we were hindered by the unavailability of a sufficiently large annotated dataset for training and testing our model. 

Instead, we use a representation of the linguistic features as presented in AutoVC~\cite{qian2019autovc}. This representation, described in the paper as the content embedding, does not require explicit phoneme annotations and can be extracted from the clean speech signal. We use the AutoVC model as a teacher network for knowledge distillation to train a network to learn the content embedding from an input mixture spectrogram. A decoder is then trained to generate vocoder features given this content embedding, which are then used for vocal synthesis. While we acknowledged in our previous work that the use of a mel-spectrogram based wavenet vocoder~\cite{shen2018natural} would lead to an improvement in synthesis quality, we believe that the ease and speed of use of vocoder features sufficiently offsets the slight degradation in audio quality for research purposes. As such, we consider the raw vocals re-synthesised by the WORLD vocoder as the upper limit for the performance of our algorithm. 


We tried our algorithm for both singing with musical mixtures and unison choir singing, which includes several singers singing at similar pitches and found the methodology to work effectively for both scenarios. In this paper, we only present the musical mixtures case, but present online examples for both use cases. We evaluate our network in terms of intelligibility of the output, isolation from the backing track and audio quality against a state-of-the-art source separation baseline via a subjective listening test. Since the output of our system is a re-synthesis of the original raw vocal track, we found the use of the standard \textit{SDR} metric \cite{fevotte2005bss_eval} to be ineffective for objective evaluation. 

 The source code for our model is available online\footnote{\url{https://github.com/pc2752/sep_content/}}, as are sound examples\footnote{\url{https://pc2752.github.io/sep_content/}}, showcasing the robustness of the model with use on real-world examples and on unison choir singing. 

\section{Related Work}
\label{sec:SOTA}


U-net~\cite{jansson2017singing} is one of the best performing algorithms for the task singing voice separation. The U-Net model takes the magnitude spectrogram of the mixture as input and applies a series of convolutional operations, resulting in a lower dimensional embedding. This lower dimensional embedding is then upsampled, following the steps of the encoder, to generate an output with the same dimensions as the input. There are skip connections between the corresponding layers of the encoder and the decoder, leading to the nomenclature of the U-Net. The output is then treated as a mask, which is applied to the input spectrogram, resulting in the desired source, in this case: the singing voice. This model assumes that the mixture is a linear sum of the individual sources and the difference between the input spectrogram and the masked input is a representation of the magnitude spectrum of the backing track. Both sources are re-synthesised using the phase of the original spectrogram.

The Unet architecture has also been applied to directly estimating the waveform~\cite{stoller2018wave}, thus avoiding the phase estimation problem. This model also estimates the \textit{stem} of the sources and follows the energy conservation assumption that the mixture is a linear sum of the individual \textit{stems}. Unlike these models, our proposed model generates a version of the raw unprocessed vocal track and not the \textit{stem}.

The work closest to ours is~\cite{biadsy2019parrotron}, which uses an end-to-end sequence-to-sequence model to extract the loudest signal from a mix of speech signal. While it is useful in a source extraction scenario, the primary application of this is voice conversion and it requires a dataset of parallel paired input-output speech utterances.

For our proposed methodology, we require a representation of linguistic information. Such representations are often used for speaker conversion systems or low resource speech synthesis systems. Algorithms such as~\cite{biadsy2019parrotron, chou2018multi, hosseini2018multi, hsu2016voice, van2017neural} use different types of content representations, learned from a clean input voice signal. For our study, we use the speaker-dependent version of AutoVC~\cite{qian2019autovc}, which is one of the most effective speaker conversion algorithms. This algorithm follows an encoder-decoder architecture and imposes restrictions on the size of the bottleneck of the encoder, thereby constraining it to learn only content based information. As shown in Equation \ref{eq:avc1}, the content encoder $E_{avc}$, takes as input the mel-spectrogram of the input speech signal, $X$, along with the speaker identity as a one-hot vector, $S$ to produce the content embedding, $C_{avc}$.
The decoder of the network, $D_{avc}$ then takes the learned content embedding, $C_{avc}$ along with the speaker identity $S$ to produce the corresponding output mel-spectrogram, $\hat{X}$, as shown in Equation \ref{eq:avc2}. 
\begin{equation}
C_{avc} = E_{avc}(X,S)
\label{eq:avc1}
\end{equation}

\begin{equation}
\hat{X} = D_{avc}(C_{avc},S)
\label{eq:avc2}
\end{equation}

The entire network is trained to minimize the reconstruction loss, $\mathcal{L}_{recon}$, shown in Equation \ref{eq:avc3}.
\begin{equation}
\mathcal{L}_{recon} = \mathbb{E} [\|\hat{X} - X\|^2_2] 
\label{eq:avc3}
\end{equation}
An additional content loss, $\mathcal{L}_{content}$, shown in Equation \ref{eq:avc4} is added to ensure that the content embedding for the output matches that of the input. 

\begin{equation}
\mathcal{L}_{content} = \mathbb{E} [\|E_{avc}(\hat{X},S) - C_{avc}\|_1] 
\label{eq:avc4}
\end{equation}

This leads to the final loss for the model, $L_{avc}$, shown in Equation \ref{eq:avc5}
\begin{equation} 
\mathcal{L}_{avc} = \mathcal{L}_{recon} + \lambda \mathcal{L}_{content}
\label{eq:avc5}
\end{equation} 

where $\lambda$ is a weight applied to the content loss, and is set to \num{1}.

\section{Methodology}
\label{sec:Methodology}

\begin{figure}
\centering
\includegraphics[width=0.5\textwidth]{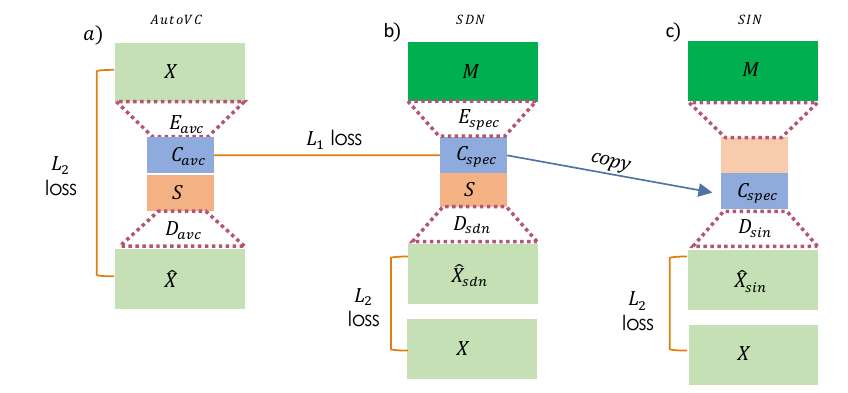}
 \caption{The three networks used in our study. Part a) shows the AutoVC~\cite{qian2019autovc} network. The content embedding learned by this network is used to train the encoder part of the second network, SDN, shown in part b). The encoder for this network is singer independent, but the decoder is singer dependent. The embedding from this encoder, along with the mixture spectrogram is passed to a third network, SIN, shown in part c). Both the encoder and decoder for this network are independent of the singer.}
 \label{fig:singer}
\end{figure}

The basic pipeline of our system is shown in Figure \ref{fig:singer}. 
We first train the AutoVC network on clean vocoder features extracted from singing voice signals to encode the content embedding as described in~\cite{qian2019autovc}. While the original AutoVC model uses the mel-spectrogram as a representation of the vocal signal, we use WORLD vocoder~\cite{morise2016world} features. The reason for this is two-fold; firstly re-synthesis from vocoder features is easier than from mel-spectrogram features and the use of WORLD vocoder~\cite{morise2016world} features allows us to disentangle linguistic content from expressive content, present in the fundamental frequency. The encoder and decoder of the network both take the singer identity as a one-hot vector, following the methodology suggested in \cite{qian2019autovc}. We tested the voice change capability of this network and found it to be effective for changing the timbre but not the expression, thereby fulfilling our criteria. The network is trained with same loss functions as used in AutoVC, shown in Equation \ref{eq:avc5}, with the exception that $X$ refers to the vocoder features pertaining to the voice signal, instead of the mel-spectrogram.

Once we have the AutoVC network trained, we use it as a teacher network for training a singer independent encoder, as shown in Figure \ref{fig:singer} b). This encoder, $E_{spec}$ has an architecture similar to $E_{avc}$ and takes the magnitude spectrogram, $M$ of the mixture signal as input and outputs a content embedding, $C_{spec}$, that matches the corresponding content embedding, $C_{avc}$. As the singer identity is not provided to the encoder, it is singer independent. The corresponding loss function is shown in Equation \ref{eq:scse}.

\begin{equation} 
\begin{aligned}
C_{spec} = E_{spec}(M)\\
\mathcal{L}_{encoder} =\mathbb{E} [\|C_{spec} - C_{avc}\|_1] 
\end{aligned}
\label{eq:scse}
\end{equation} 

The decoder part of the model, $D_{sdn}$, takes as input the content embedding, $C_{spec}$, along with singer identity as a one-hot vector, $S$, and outputs the clean vocoder features, $\hat{X}_{sdn}$ of the corresponding vocal track. Equation \ref{eq:scsd} shows the loss function for this network. The architecture for the decoder is the same as $D_{avc}$. This system is henceforth referred to as the singer dependent network (SDN).

\begin{equation} 
\begin{aligned}
\hat{X}_{sdn} = D_{sdn}(C_{spec}, S)\\
\mathcal{L}_{D_{sdn}} = \mathbb{E} [\|\hat{X}_{sdn} - X\|^2_2] 
\end{aligned}
\label{eq:scsd}
\end{equation} 

Finally, we train a singer independent decoder, $D_{sin}$, which takes the content embedding, $C_{spec}$, of the previously trained singer independent encoder along with the mixture spectrogram as input and outputs the corresponding vocoder features, $\hat{X}_{sin}$. Figure \ref{fig:singer} c) shows the data-flow used in this model, while the loss function is represented by Equation \ref{eq:scmd}. This system is henceforth referred to as the singer independent network (SIN).

\begin{equation} 
\begin{aligned}
\hat{X}_{spec} = D_{sin}(C_{spec}, M)\\
\mathcal{L}_{D_{sin}} = \mathbb{E} [\|\hat{X}_{sin} - X\|^2_2] 
\end{aligned}
\label{eq:scmd}
\end{equation} 

While the training part of the networks is inter-related, the feed-forward prediction of an individual network is independent of the other networks

For fundamental frequency estimation, we use a network similar to the one proposed in~\cite{janssonjoint}, trained on the same data as the other networks. We use this prediction along with the vocoder features to synthesise the audio signal. We tried both the discrete representation of the fundamental frequency as described in \cite{janssonjoint} and a continuous representation, normalised to the range \num{0} to \num{1} as used in \cite{chandna2019vocoder} and found that while the discrete representation leads to slightly higher accuracy in the output, the continuous representation produces a pitch contour perceptually more suitable for synthesis of the signal.



\section{Experiments}
\label{sec:eval}
\subsection{Dataset}
\label{sec:dataset}
We use a proprietary dataset for training the model and the MedleyDB dataset~\cite{bittner2014medleydb} for evaluation of the model. The training set consists of \num{205} songs by \num{45} distinct male and female singers, with a total of around \num{12} hours of data. the songs are mostly pop songs in the English and Japanese languages. 
We use \SI{90}{\percent} of the proprietary dataset for training and \SI{10}{\percent} for validation, which we use for early stopping during training of the model. We have access to the raw vocal track in this training set as well as annotation of the singers, which makes it ideal for our proposed model.

For testing, we use the the MedleyDB dataset, which contains \num{122} songs. The raw audio tracks and the mixing \textit{stems} for which are present. We use the raw audio vocal tracks of \num{6} of the songs for computing the vocoder features, which are used to re-synthesise the singing track and are used as a reference during evaluation. To the best of our knowledge, there is no overlap amongst the singers in the training set and the singers present in MedleyDB. Therefore the use of this dataset for evaluation makes sense as we are using both songs and singers not seen by the model during training.

The audio files were all downsampled to a sampling rate of \SI{32}{\kilo\hertz}. The short time fourier transform (STFT) was calculated with a Hanning window of size \num{1024}. Both the STFT and the vocoder features were calculated with a hoptime of \SI{5}{\milli\second}.
We use dimensionality reduction for the vocoder features~\cite{blaauw2017neural, tokuda1994mel, chandna2019vocoder}, leading to \num{64} features.
\subsection{Training}
As suggested in the AutoVC~\cite{qian2019autovc} model, we use the Adam~\cite{kingma2014adam} optimizer for training the networks. A batch size of \num{30} is used with an input of length \SI{640}{\milli\second}, randomly sampled from the tracks. We augment the input data by applying variable gains to the vocal and backing track spectrograms while creating the mixture during training. 


\subsection{Evaluation Methodology}
There are three aspects of the signal predicted by our model to be evaluated; the intelligibility of the produced signal, isolation from the backing track and the quality of the signal. The prediction of our model is different from that made by most state-of-the-art source separation models in that our model predicts the raw unprocessed vocals whereas other source separation models predict the processed \textit{stem}. Thus, a direct comparison with other models via objective measures like those presented in \textit{bss$\_$eval}~\cite{fevotte2005bss_eval} is not feasible. Instead we use a subjective listening test for evaluation~\cite{kraft2014beaqlejs}. We evaluate both the singer dependent network,  SDN and the singer independent network, SIN, against a baseline~\cite{jansson2017singing}, referred to as UNET and against our previously proposed model~\cite{chandna2019vocoder}, SS. We adjusted the U-Net architecture to account for the change in the sampling frequency and the window size used in our model. All four models were trained and evaluated on the same datasets as described above. For subjective evaluation, we used an online AB listening test, wherein participants were asked to pick between two examples on the criterion of intelligibility, source isolation and audio quality. Each criteria had \num{15} questions with \num{5} for each pair.
We also use Mel Cepstral Distortion (MCD) compared to the raw vocal signal, as shown in table \ref{table:models} as an objective measure for overall audio quality of the vocal synthesis.

\begin{table}[h]
\centering

\begin{tabular}{l c c c c }

Model & MCD MEAN (dB) & MCD STD (dB) \\ \hline
SS & 7.39 & 1.25 \\
SDN & 7.55 & 0.63 \\
SIN & 6.45 & 0.75\\
UNET  & 6.58 & 1.88\\

\end{tabular}
\caption{The Mel Cepstral Distortion (MCD) metric in dB. It can be observed that the singer independent network, SIN, is comparable in quality to UNET.}
\label{table:models}
\end{table}




\subsection{Subjective Evaluation}
\begin{figure}[H]
\centering
\includegraphics[width=0.5\textwidth]{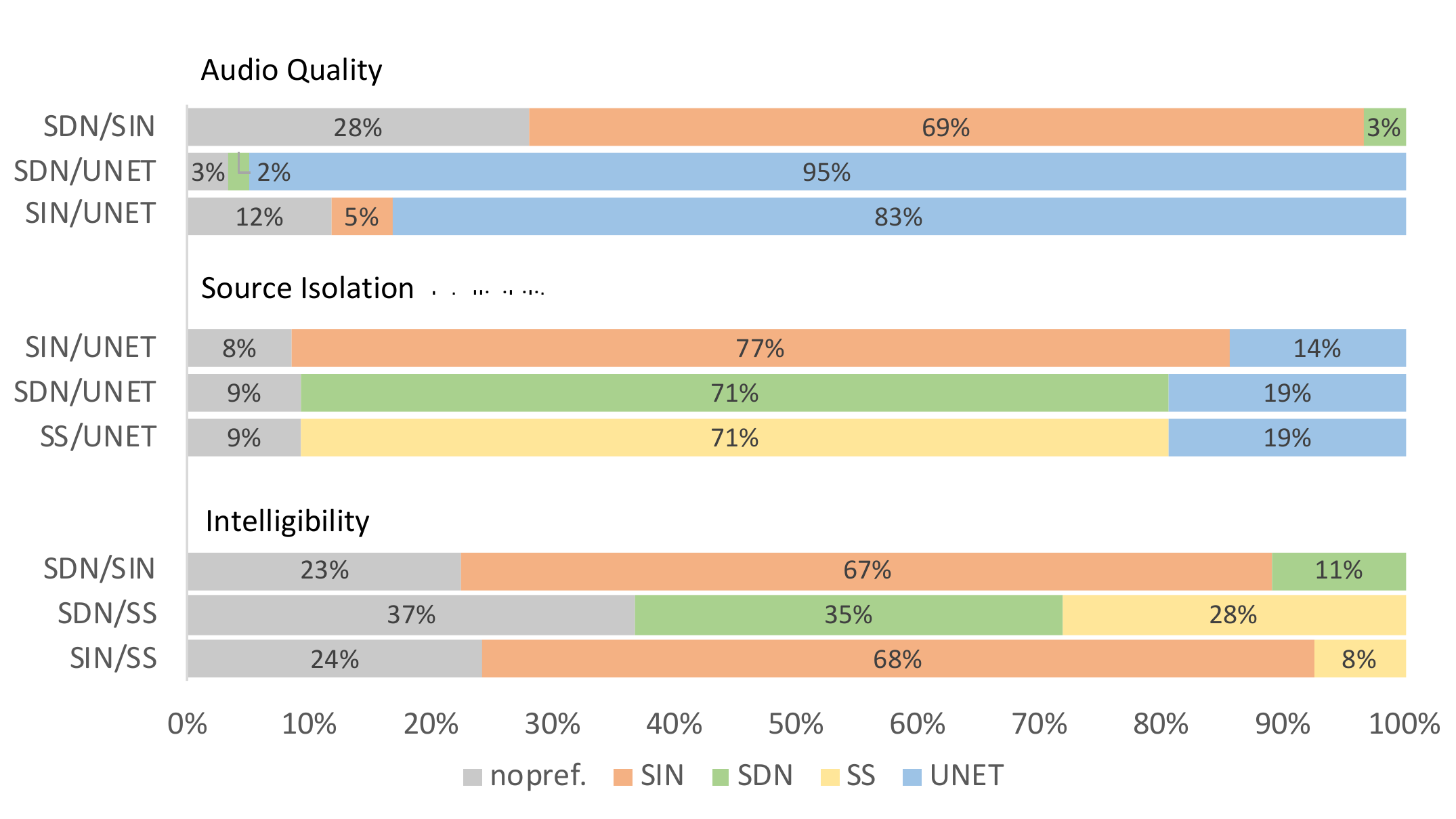}
 \caption{Results of the listening test. The vocoder based models outperform UNET~\cite{jansson2017singing} in terms of source isolation, but lag behind in terms of audio quality. The proposed models, SDN and SIN, are ranked higher than our previous model, SS~\cite{chandna2019vocoder} in terms of intelligibility, showing an improvement over the baseline.}
 \label{fig:intell}
\end{figure}

There were \num{25} participants in our listening test, from various countries. All participants claimed proficiency in the English language, that was used in our survey and \num{18} had previous musical training. The results of the listening test are shown in Figure \ref{fig:intell}. Audio quality and intelligibility is observed to be lower for SDN than SIN, this is expected, since the SDN model produces an output given only the content embedding and the singer identity whereas the decoder of the SIN model has access to more information via the magnitude spectrogram of the input. 

We can see that both the SDN and SIN are ranked higher on intelligibility than SS, thus showing that the content encoder is able to effectively learn the underlying linguistic features from the given input spectrogram in a singer independent manner, even for singers not seen during training and for a mixed and processed vocal track. 
We also observe that all three vocoder based models outperform the mask based model, UNET on isolating the vocal signal from the mixture. This follows from our previous observation~\cite{chandna2019vocoder} and is due to the approach that we follow of re-synthesising the vocal signal from the mixture rather than using a mask over the input signal. 
The models still lag behind on audio quality. This can partly be attributed to the degradation by the use of vocoder features and to the errors in estimation of the fundamental frequency, which is done via an external model. However, it can be seen that the proposed SIN model outperforms the baseline of the SS model, which can be explained to the difference in architecture and the additional information from the content embedding that the network receives.

\section{Conclusions}
\label{sec:conclusions}
We present a methodology to extract the raw unprocessed vocal signal from a musical mixture based on a linguistic feature extractor. We show that using knowledge distillation from a teacher network, we are able to train an encoder to extract linguistic features from a mixture spectrogram and use these features to synthesise the underlying vocal signal. The model is able to synthesise intelligible vocals, even for an unseen singer, without interference from the backing track. Using a decoder that takes the mixture spectrogram as an input instead of the singer identity leads to a significant improvement in audio quality and intelligibility over the baseline vocoder based model that we presented earlier. 
An additional application of our system is for isolating voices in unison singing, a task that has not been tackled so far by traditional source separation methodologies. 



\vfill\pagebreak

\bibliographystyle{IEEEbib}
{\small
\bibliography{strings,refs}}

\end{document}